\documentstyle[aps,epsf,floats]{revtex}

\begin{document}
\newcommand{\eg}{\mbox{e.\ g.\ }}
\newcommand{\ie}{\mbox{i.\ e.\ }}
%\draft
\twocolumn[\hsize\textwidth\columnwidth\hsize\csname@twocolumnfalse%
\endcsname

\title{Evidence For Heavy Hole and Light Hole Current Separation In P-type Resonant Tunneling Diodes With Prewells}
\author{R.\ M.\  Lewis,  H.\ P.\  Wei}
\address{Department of Physics, Indiana University, Bloomington, Indiana 47405, U.\ S.\ A.\ \\ }
\author{S.Y. Lin and J.\ F. Klem}
\address{Sandia National Laboratories, Albuquerque, New Mexico 87185 U.\ S.\ A.\ \\}
\date{\today}
\maketitle
%
%\draft
%
\begin{abstract}
We investigate the transport of holes through $AlAs/In_{.10}Ga_{.90}As$ resonant tunneling diodes which utilize $In_xGa_{1-x}As$ prewells in the emitter with $x=0,.10,$ and $.20$.  The data show an increase in peak current and bias at resonance and a concurrent increase in the peak--to--valley ratio with increasing x.  We explain this enhancement in tunneling as due to confinement (or localization) of charges in the prewell and the formation of direct heavy(light) hole to heavy(light) hole conduction channels as a consequence.
\end{abstract}
\pacs{PACS numbers;}]

\narrowtext
\smallskip
	Charge transport through resonant tunneling diodes (RTD) at low bias is completely determined by the states in the quantum well.  The peaks in the tunneling current versus bias voltage (I--V) data correspond to the Fermi energy in the emitter lining up with the energy of the state in the well\cite{changesakitsu}.  Therefore, changes to the structure that affect how the energy levels align will show up in the I--V data.  For instance, the quantized states in a wide quantum well have lower energies than those in a narrow well and so the resonant peaks in I--V occur at lower bias.  Similarly, by modifying the structure before the emitter barrier and leaving everything else unchanged, it is possible to explore the effects of different emitter configurations.
	
In this letter we report on data from three p-type GaAs/AlAs RTD with In$_{x}$Ga$_{1-x}$As prewells of varying depth.  Because the only parameter which changes in our structures is the depth of the prewell, we are able to quantify its effects on transport in p-type RTD. Prewells were first utilized by Lee and Harris \cite{leeandharris} and by Reichert {\it et al.} \cite{riechert} both of whom saw increased quality as defined by the peak-to-valley ratio of their devices.  More recently, Boykin {\it et al.} \cite{boykin} have compared calculation and experiments for a systematical study of prewells of different widths.  However, all these efforts have focused on n-type RTD.  In p-type devices, the situation is more complicated because heavy holes (HH) and light holes (LH) are degenerate in the valence band at  wavevector {\bf k}$=0$.  However, when confined in a quantum well, it is possible to distinguish the HH and LH states \cite{mendez}.  Because HH and LH carry different quantum numbers \cite{singh}, selecting carrier type in the emitter offers a way to increase the tunneling current.  The data we present here show an improvement in the quality of the devices and indicate that selection of carrier type in the emitter is possible.

Our samples were grown by molecular beam epitaxy on p-type $GaAs$ substrates.  The active region of our devices is formed by: 3000 $\AA$ of $GaAs$, $Be$ doped to $2 \times 10^{18} \ cm^{-3}$; 1000 $\AA$ of $GaAs$, $Be$ doped to $1 \times 10^{18} \ cm^{-3}$; 1000 $\AA$ of $GaAs$, $Be$ doped to $5 \times 10^{17} \ cm^{-3}$; 50 $\AA$ of $GaAs$ (undoped spacer); 70 $\AA$ of $In_{x}Ga_{1-x}As$ undoped where $x=0, .10$, or $.20$ (prewell); 60 $\AA$ of $AlAs$ undoped (emitter barrier); 34 $\AA$ of $In_{.10}Ga_{.90}As$ undoped (well);50 $\AA$ of $AlAs$ (collector barrier); 500 $\AA$ of $GaAs$ undoped (spacer layer); 1000 $\AA$ of $GaAs$, $Be$ doped to $5 \times 10^{17} \ cm^{-3}$;  1000 $\AA$ of $GaAs$, $Be$ doped to $1 \times 10^{18} \ cm^{-3}$; 3000 $\AA$ of $GaAs$, $Be$ doped to $2 \times 10^{18} \ cm^{-3}$.  Note that the emitter barrier is thicker than the collector barrier to inhibit charge build-up in the well under forward bias.  Also, the quantum well includes 10 percent indium composition, creating some built in strain.  The principle feature of interest is the presence of a prewell in the emitters of sample B (10 percent indium) and C (20 percent indium), while sample A (no indium) is without a prewell.

The samples were defined as $50\mu m \times 50 \mu m$ mesas using photolithography.  A thin AuZn alloy layer approximately 1000 $\AA$ thick was evaporated onto these squares for use as contact pads and as a mask for wet etching.  A polyimide \cite{polyimide} insulating layer approximately $1 \mu m$ thick was spun onto the surface and dried at 90$^0$ C for about 1 hour.  Contact windows measuring $20 \mu m \times 20 \mu m$ were opened in the polyimide layer.  Full curing of the insulating layer was then accomplished by baking at $T > 185^0$C for at least 45 minutes.  The samples were annealed for 20 seconds at 400$^0$C in a reducing atmosphere to ensure ohmic contacts.  Transport measurements were performed at 4.2K in helium using AC lock-in technique and a DC biasing circuit. 

\begin{figure}[tb]
\begin{center}
\leavevmode
\epsfxsize \columnwidth
\epsffile{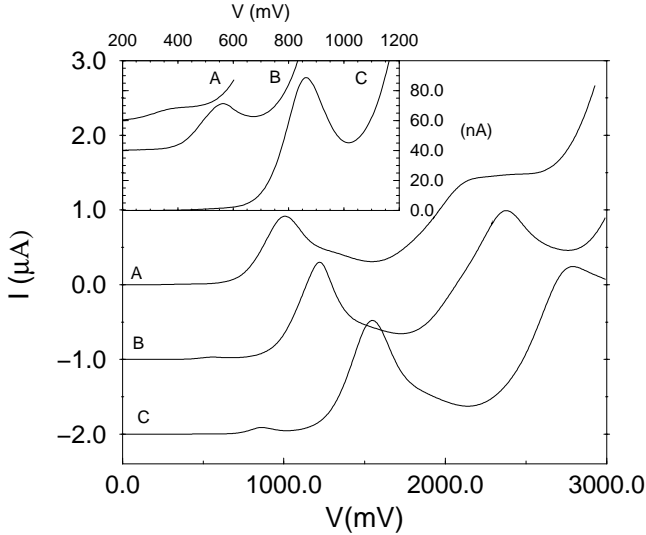}
\caption{I versus V for sample A, B, and C. Data for samples B and C has been shifted for clarity or presentation.  The inset shows just the low bias region around the HH resonance for the three samples, again shifted.  }
\label{fig1.}
\end{center}
\end{figure}

Figure 1 plots current I versus forward voltage bias V for all three samples, over the full resonant tunneling range.  Three resonances appear in each of these curves which are best understood by considering the simplest case, sample A.  The data for sample A shows its principle features at $V=365$mV, $V=1007$mV, and $V=2107$mV.  We can determine the peak position to within 10mV by looking at measured derivatives of the I--V curve.  The secondary features which show up at $V  \approx 1325 $mV and on either side of the 3rd principle resonance are related to their respective resonances and are not important for this discussion.

	We estimate the mass of the states in the quantum well in sample A by modeling the system under applied bias.  Our model follows Goldman {\it et al.}\cite{goldman}, except that we assume no charge accumulation in the well and we treat the emitter as a shallow well, with no charge build up.  It can be shown that the overall energies of the states in the well are not very sensitive to the exact emitter band bending profile.  Using the peak voltages, we find the effective mass of the state corresponding to the first peak in the I--V data to be $0.24 \pm .02$ $m_e$ where $m_e$ is the free electron mass.  The state associated with the second peak has a mass of about 0.096 $m_e \pm .002$.  The error bars are obtained by varying parameters within the model.  In bulk $In_{.10}Ga_{.90}As$, calculating the masses using the Luttinger parameters gives $m_{hh}=0.32 m_e$ and $m_{lh}=.073 m_e$ \cite{note3}, suggesting that the first resonance in the I--V data is due to conduction through the HH$_0$ state in the well, and the second through a LH$_0$ state.  Calculations of the bound states in a square well reinforce these identifications, showing that one should expect two HH and one LH states. 

%*** Experimental values give $m_{hh}=0.45$ and $m_{lh}=0.09$ for GaAs(cite reithmair maybe).***  Of course, confinement changes the band structure, shifting the energies\cite{foreman}.

The lower two traces in figure 1 are the I--V data from samples B and C. These are nearly identical to the data from sample A, with two strong resonances and one much weaker resonance visible in the forward bias.  All prominent features in curve A shift systematically to higher bias in curves B and C.  More explicitly, the HH$_0$ peak shifts from 365mV in A, to 564mV in B, and to 861mV in C and the LH$_0$ peak occurs at 1007mV in A, 1222mV in B, and 1550mV in C.  This shift indicates that the emitter states are settling into a well whose depth changes from sample to sample.  As the emitter state's energy changes, more bias is needed to reach resonance.  Note that the peak current for the LH$_0$ resonance is about 0.90$\mu$A in sample A.  In sample C this has grown to about 1.5$\mu$A.  The inset of figure 1 replots the I--V data around the first resonance to better show the increase in strength of the HH$_0$ current peak.  Note that the scale is now nA and that the data for sample A and B are shifted up by 60nA and 40nA.  Current at $V=365$mV is just 7.5nA in sample A, but has grown to 86nA in sample C, an eleven-fold increase which dwarfs the factor of 2 change in the LH$_0$ resonance.
	In terms of peak-to-valley ratios (PVR), for the HH$_0$ resonance these current numbers equate to 0.7 for sample A, 1.4 for sample B, and 2.0 for sample C.  For the LH$_0$ resonance the PVR's are 3.0 for sample A, 3.8 for sample B, and 4.0 for sample C.
 
\begin{figure}[tb]
\begin{center}
\leavevmode
\epsfxsize \columnwidth
\epsffile{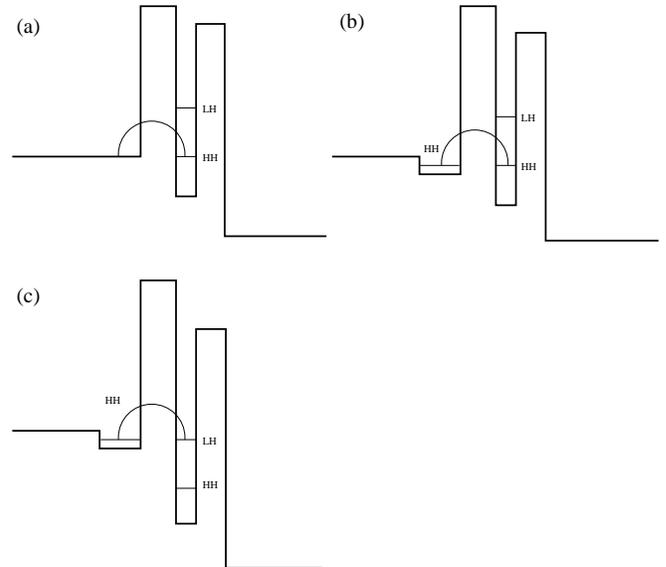}
\caption{Three schematic flat-band models to illustrate the tunneling processes in our samples.}
\label{fig2.}
\end{center}
\end{figure}

To help explain the changes induced by the prewell consider Figure 2, a flat-band schematic of the resonant tunneling process where for simplicity all the bias is dropped across the barriers, and hole energies {\it rise} vertically.  In (a), a device with no emitter prewell is shown.  Tunneling is allowed to take place when a carrier in the valence band arrives at the emitter barrier and the applied bias is sufficient that energy levels on each side of the barrier line up.   Panel (b) shows a similar set of conditions, but now a prewell is included in the device.  The effect of the prewell will be to localize carriers near the barrier at an energy slightly below the valence band.  The two direct consequences seen in the I--V traces are the moving the resonances to higher bias and the increase of peak current and PVR.  The the Tsu-Esaki equation \cite{tsu-esaki} states that the tunneling current is proportional to the overlap integral of the densities of states on each side of the emitter barrier summed over {\bf k} parallel to the barrier.  The overlap integral can be enhanced through localization of the carriers as in (b) for 2 reasons.  First, the prewell ensures a {\it greater number} of charges near the barrier.  Second, it ensures these charges are {\it near} the barrier.  Our modeling of samples B and C at resonance shows that more charge is being localized at the emitter barriers in these devices than in sample A, corroborating these speculations.

	The picture developed so far explains every aspect of the data except the discrepancy between the enhancement of the HH and LH peaks.  Our samples have quite shallow prewells, so HH are much more strongly localized in the emitter than LH.  In (b) of figure 2, a HH in the prewell tunnels into the HH ground state of the quantum well.  Of course, LH from the valence band could also contribute to the tunneling current, but the data in figure 1 indicate that the HH peak strength is strongly affected by the depth of the prewell and so presumable by the number HH in the prewell.  In (c) of figure 2, a HH in the prewell tunnels to the LH state in the well.  But, our data do not show a strong increase the the LH peak strength as the prewell is deepened.  We speculated that this is because HH in the prewell do not contribute to the current at the LH resonance.  In fact, our data support the idea that a HH to HH is an allowed process, but HH to LH is not and consequently that carrier type, LH or HH, is preserved during tunneling.

 The last point requires some clarification.  In particular, Wessel and Altarelli \cite{wesselandaltarelli} have suggested that strong mixing exists between the incoming LH and HH and hence both contribute to the current at each resonance.  However Ji {\it et al.} \cite{Ji} measured the strain induced splitting of the valence band caused by indium and Lin {\it et al.} have shown that some biaxial compressive strain in the quantum well (through indium incorporation) extends the independent character of the LH and HH subbands to finite wave vectors\cite{sylin}.  We note that both the quantum well and prewell of our samples contain approximately 10 percent indium, and thus we can assume the HH and LH subbands experience minimal or no mixing in the barriers.

	In summary, we have demonstrated the effects of prewells in p-type $GaAs/AlAs$ RTD's.  For small bias, we find the effects of a prewell are an enhanced localization of charge in the emitter and an increase in the PVR of all the resonances, but most notably, the HH$_0$ peak.  To explain this extra increased strength in the tunneling current, we propose that the prewell localizes HH preferentially in the emitter enhancing the overlap of carriers in the emitter with the well states.  A full explanation of the data requires that holes tunnel with higher probability into states of the same mass (heavy or light) and hence that tunneling prefers to conserve the mass state.  It should be possible to exploit this property in electronic devices.

We acknowledge stimulating discussions with Alex Zaslavsky, Bill Schaich, Tomas Jungwirth, and Mike Jaeger. This work is supported by grant number DMR9311091 at Indiana University and at Sandia National Laboratories through the DOE.  Sandia is a multiprogram laboratory operated by Sandia Corporation, a Lockheed Martin Company, for the United States Department of Energy.


\begin{thebibliography}{}

\bibitem{changesakitsu}L.\ L.\ Chang, L.\ Esaki, and R. Tsu, Appl.\ Phys.\ Lett.\ {\bf 24}, 593 (1974).

\bibitem{leeandharris}S.\ Lee, and J.\ S.\ Harris, Jr.\, IEEE Trans.\ Electronic Devices, {\bf 36}, 2619 (1989).

\bibitem{riechert} H.\ Riechert, D.\ Bernklau, J.\ P.\ Reithmaier, and R.\ D.\ Schnell, Electron.\ Lett.\, {\bf 26}, 341 (1990).

\bibitem{boykin}Timothy B.\ Boykin, R.\ Chris Bowen, Gerhard Klimeck, and Kevin L.\ Lear,  Appl. Phys. Lett. 75, 1302 (1999).

\bibitem{mendez}E.\ E.\ Mendez, W.\ I.\ Wang,B.\ Ricco, and L.\ Esaki Appl.\ Phys.\ Lett.\  {\bf 47}, 415 (1985).

\bibitem{singh}J.\ Singh, {\it Physics of Semiconductors and Their Heterostructures},( Magraw Hill, 1993) p 145.

\bibitem{polyimide}We use Pyralin PI 2555 from DuPont Company.

\bibitem{goldman}V.\ J.\ Goldman, D.\ C.\ Tsui, and J.\ E.\ Cunningham, Phys.\ Rev.\ B {\bf 35}, 9387 (1987).

\bibitem{note3}For GaAs, $\gamma_1=7.1+13.3x$, $\gamma_2=2.02+6.28x$ where $x$ is the indium fraction. Hole masses are calculated using $m_{hh}=m_e/(\gamma_1 -2\gamma_2)$, $m_{lh}=m_e/(\gamma_1 +2\gamma_2)$. Luttinger parameters from N.\ Binggeli and A.\ Baldereshi, Phys.\ Rev.\ B {\bf 43}, 14734 (1991).

\bibitem{tsu-esaki}R.\ Tsu and L.\ Esaki Appl. Phys. Lett. {\bf 22}, 562 (1973).

%\bibitem{luryi}S.\ Luryi, Appl.\ Phys.\ Lett.\ {\bf 47}, 490 (1985).

\bibitem{wesselandaltarelli}R.\ Wessel and M. Altarelli, Phys. Rev. B {\bf 39}, 12802 (1989). 

\bibitem{Ji}G.\ Ji, D.\ Huang, U.\ K.\ Reddy, T.\ S.\ Henderson, R.\ Houdre, and H.\ Morkoc, J.\ Appl.\ Phys.\ {\bf 62}, 3366 (1987).

%\bibitem{jpreithmaier}J.\-P.\ Reithmaier {\it et al.}, Appl.\ Phys.\ Lett. {\bf 56}, 536 (1990).

\bibitem{sylin}S.\ Y.\ Lin, A.\ Zaslavsky, k.\ Hirakawa, and D.\ C.\ Tsui,  Appl. Phys. Lett. 60, 601 (1992).

%\bibitem{zaslavsky1}A.\ Zaslavsky {\it et al}, Phys.\ Rev.\ B {\bf 47}, 16036 (1993).

%\bibitem{hendrofer} G.\ Hendrofer, Semicond.\ Sci.\ Technol.\ {\bf 9}, 1999 (1994).

%\bibitem{mendez2}E.\ E.\ Mendez, W.\ I.\ Wang, B.\ Ricco,and L.\ Esaki, Appl.\ Phys.\ Lett.\, {\bf 47}, 415 (1985).

%\bibitem{sabrown} S.\ A.\ Brown {\it et al.}, PRB {\bf 56}, 1967 (1997).

%\bibitem{tafisher} T.\ A.\ Fisher {\it et al.}, PRB {\bf 50}, 18469 (1994).

%\bibitem{foreman}Bradley Foreman, Phys.\ Rev.\ B {\bf 49} 1757, (1994).

%\bibitem{steve}R.\ E.\ Prange and S.\ M.\ Girvin, {\it The Quantum Hall Effect} (Springer--Verlag, New York, 1986).

\end{thebibliography}
\end{document}